\documentstyle[12pt]{article}
\begin{document}
\def\abs#1{\left|{#1}\right|}
\def\edet{{\det}'}
\def\mt{{\ifmmode\td M_t\else $\td M_t$\fi}}
\let\nn=\nonumber
\let\td=\tilde
\let\CR =\cr
\def\fr#1.#2.{{#1\over #2}}
\let\FR =\fr
\let\dg=\dagger
\def\yu{\ifmmode Y_u^{\phantom{\dg}}\else$Y_u^{\phantom{\dg}}$\fi}
\def\yd{\ifmmode Y_d^{\phantom{\dg}}\else$Y_d^{\phantom{\dg}}$\fi}
\def\ye{\ifmmode Y_e^{\phantom{\dg}}\else$Y_e^{\phantom{\dg}}$\fi}
\def\hyu{\ifmmode Y_u^\dg\else$Y_u^\dg$\fi}
\def\hyd{\ifmmode Y_d^\dg\else$Y_d^\dg$\fi}
\def\hye{\ifmmode Y_e^\dg\else$Y_e^\dg$\fi}
\def\cs#1{{\ifmmode{\cal S}_{#1}\else${\cal S}_{#1}$\fi}}
\def\mg{{\ifmmode M_{GUT}\else $M_{GUT}$\fi}}
\def\G{\Gamma^{\vphantom{*}}}
\def\cjkl{C_{ijkl}}
\let\lam=\lambda
\def\cqqh{\hat{c}_{\mathstrut qq}{}}
\def\cqlh{\hat{c}_{\mathstrut ql}{}}
\def\cqlw{\check{c}_{\mathstrut ql}{}}
\def\cudh{\hat{c}_{\mathstrut ud}^*{}}
\def\cueh{\hat{c}_{\mathstrut ue}^*{}}
\let\ep=\epsilon
\def\abg{{\alpha \beta \gamma}}

\thispagestyle{empty}
\noindent
\begin{flushright}
        OSHTPY-HEP-T-00-012\\
        July 17, 2000
\end{flushright}

\vspace{1cm}
\begin{center}
  \begin{Large}
  \begin{bf}
       SUSY GUTs under Siege : Proton Decay 
  \end{bf}
 \end{Large}
\end{center}
  \vspace{1cm}
    \begin{center}
    Radovan Derm\' \i \v sek, Arash Mafi and Stuart Raby  \\
      \vspace{0.3cm}
\begin{it}
Department of Physics,
The Ohio State University,
Columbus, Ohio  43210
\end{it}
  \end{center}
  \vspace{1cm}
\centerline{\bf Abstract}
\begin{quotation}
\noindent
SO(10) supersymmetric grand unified theories [SUSY GUTs] provide a beautiful 
framework for physics beyond the standard model.   Experimental 
measurements of the three gauge couplings are consistent with 
unification at a scale $M_G \sim 3 \times 10^{16}$ GeV.  In addition
predictive models for fermion masses and mixing angles have been
found which fit the low energy data, including the recent data for neutrino
oscillations.   SO(10) boundary conditions can be tested via the spectrum
of superparticles.  The simplest models also predict neutron and 
proton decay rates.   In this paper we discuss nucleon decay rates and obtain
reasonable upper bounds.   A clear picture of the allowed SUSY spectra
as constrained by nucleon decay is presented.

\end{quotation}
\vfill\eject

\section{Introduction}
 
The standard model is unlikely to be a fundamental theory; it contains
19 arbitrary parameters, 13 of which are the charged fermion masses and mixing
angles.  At least six more parameters are needed to describe neutrino masses and 
mixing (for three active neutrinos).   Supersymmetric [SUSY] grand unified theories [GUTs] provide a beautiful framework for understanding many of the outstanding problems of the standard model\cite{physicstoday}.  \\

For this framework to be accepted as a description of Nature, the three pillars of SUSY GUTs must be verified.  These are
\begin{itemize}
\item{[I]} Gauge and Yukawa Coupling Unification\cite{drw,yukawaunification,predictivemodels};
\item{[II]} Observable Superparticle Spectrum\cite{so10boundaryconditions};
\item{[III]} Nucleon Decay\cite{nucleondecay}
\end{itemize}

Of these only (I) has been verified.   (II) must await RunII at the Fermilab
Tevatron or the LHC and (III) is in danger of being observed or excluded by
SuperKamiokande and Soudan II.  It is this last feature which is the main subject of this letter.  In the next section we review the steps
in the calculation of nucleon decay rates.  

\section{Nucleon Decay}

In SUSY GUTs, dimension five baryon and lepton number violating operators resulting from the exchange of color triplet Higgsinos dominate; suppressed by one power of an effective color triplet mass \mt.   Unlike proton decay mediated by gauge boson exchange, the value of $\mt$  is a free parameter.  It is only constrained by requiring perturbative threshold corrections at $M_G$.  This constraint, in conjunction with nucleon decay bounds, however, has been used to rule out simple SU(5) SUSY GUTs\cite{su5}.  SO(10) on the other hand, has escaped exclusion, but it is now under siege. 

 In equation 1, the Higgs doublets $H_u \;(H_d)$ and the color triplets  $T \; (\overline T)$ are contained in a single 10 dimensional representation $10_H$ of SO(10).  The couplings of the color triplet Higgs fields to quarks and leptons are given in terms of Yukawa-like $3 \times 3$ complex matrices $c_{qq}$, $c_{ud}$, $c_{ue}$, $c_{ql}$.  These are related by SO(10) to the Yukawa matrices $Y_u$, $Y_d$, $Y_e$ (eqn. 1).\footnote{For more details on notation, see \cite{lucas1}.}  In a predictive SUSY GUT the arbitrary parameters in the Yukawa matrices, defined at the GUT scale, are fixed when fitting charged fermion masses and mixing angles at low energies. Hence the parameters in the flavor matrices $c_{qq}$, $c_{ud}$, $c_{ue}$, $c_{ql}$ are also fixed once charged fermion masses are fit.\footnote{ In our analysis the Yukawa matrices and gauge couplings are renormalized from $M_G$ to $M_Z$ using two loop SUSY renormalization group equations [RGEs].  We then use a global $\chi^2$ analysis to fit the data\cite{bcrw}.  This analysis self consistently checks for electroweak symmetry breaking and includes the relevant one loop threshold corrections at $M_Z$.} 
\begin{eqnarray} 
 &H_u Q Y_u \overline U  + H_d Q Y_d \overline D + H_d L Y_e \overline E & \nn \\ & + Q {1\over 2}c_{qq} Q T+Q c_{ql} L \overline T+\overline U c_{ud}
\overline D  \overline T+\overline U c_{ue}
\overline E T & 
\end{eqnarray}

Below the GUT scale the color triplets are integrated out of the theory giving the dimension five operators (eqn. 2).  All dimensionless (dimensionful) parameters are then renormalized to $M_Z$.$^2$ We use universal squark and slepton masses ($m_0$), gaugino masses ($M_{(1/2)}$) and non-universal Higgs masses ($m_{H_u}, \; m_{H_d}$) at $M_G$. 

\begin{eqnarray} 
 &H_u Q Y_u \overline U + H_d Q Y_d \overline D + H_d L Y_e \overline E \ & \nn \\ & + {1\over{\mt}^{\phantom{(}}}Q {1\over 2}c_{qq} Q \ Q c_{ql}
L+{1\over{\mt}^{\phantom{(}}}\overline U c_{ud} \overline D 
\ \overline U c_{ue} \overline E &  
\end{eqnarray}

In the effective theory below $M_Z$ the coefficients of the effective (dimension six) four fermi baryon and lepton number violating operators are determined. \footnote{In a more detailed calculation it may be appropriate to have a hierarchy of effective field theories to take into account the hierarchy of SUSY particles between 100 GeV and 3 TeV.}  These are obtained via one loop graphs with squark, slepton and gaugino intermediate lines.   In general there are LLLL, LLRR and RRRR operators generated via gluino and chargino exchanges.\footnote{We have not included the contribution of neutralino loops.  For more details, see for example \cite{lucas1}.}

We then renormalize the four fermi operators from $M_Z$ to 1 GeV using 
QCD with the multiplicative factor $A_3$\cite{rge}.\footnote{Note, $A_3$ is different than $A_L$ which appears in \cite{AL} and is used in many other works on proton decay.   This is because $A_L$ takes into account two different effects, the QCD running of the dimension 6 operators from $M_Z$ to 1 GeV and the running of the quark masses from low energies to $M_Z$ in order to use the correct Yukawa couplings at the weak scale.  In our analysis $A_3$, including only the QCD running of the dimension 6 operator, is the appropriate factor to use, since we are already using the Yukawa couplings evaluated at $M_Z$.  The general solution to the RGE for the coefficient $C(\mu)$ of the dimension six operator is given by  $ C(\mu) = C(\mu_0) (\frac{\alpha_s(\mu)}{\alpha_s(\mu_0)})^{(2/b_0)}$
where $b_0 = 11 - \frac{2}{3} n_{flavors}$.  Note also that analytically we have
the relation $A_L =  A_3^{-3}$.  Given the numerical value we find for $A_3 = 1.32$, we then obtain the value of $A_L = 0.43$.  For some reason, this differs from the value of $A_L = 0.22$\cite{AL} used in previous works.  We believe this must be a numerical error.}

\begin{eqnarray}
A_3 = & \left(\frac{\alpha_s(1 {\rm GeV})}{\alpha_s({\rm m_c})}\right)^{(2/9)}
\left(\frac{\alpha_s({\rm m_c})}{\alpha_s({\rm m_b})}\right)^{(6/25)} \left(\frac{\alpha_s({\rm m_b})}{\alpha_s({\rm M_Z})}\right)^{(6/23)} & = 1.32
\end{eqnarray}

The final step is to evaluate the matrix elements of these four fermi operators between a nucleon and the lepton + meson final state.   This requires lattice gauge theory calculations and usually chiral Lagrangian analysis\cite{chadha,beta}. The decay rates depend significantly on the chiral Lagrangian factors $\alpha_{lat}$ and
$\beta_{lat}$ where $$\beta_{lat} U({\bf k}) =\ep_\abg \smash{<}0 | (u^\alpha d^\beta)
u^\gamma |{\rm proton}({\bf k})\smash{>} ,$$  $$\alpha_{lat} U({\bf k}) =\ep_\abg
\smash{<}0 |(\overline{\vphantom{d}u}^{*\,\alpha}
\overline d^{*\,\beta}) u^\gamma |{\rm proton}({\bf k})\smash{>}$$ and $U({\bf
k})$ is the left handed component of the proton's wavefunction. 

We finally obtain the general amplitude for Higgsino mediated nucleon decay given schematically by the formula
\begin{eqnarray}
T \propto & A_3 \; ({\rm FF})  \; ({\rm LF}) \; \mt^{-1} \;  \beta_{lat} & 
\end{eqnarray}
where 
\begin{itemize}
\item  $FF$ is a {\bf Flavor Factor} depending on Yukawa, gauge couplings and $c_{qq}\; c_{ql}$, $c_{ud} \; c_{ue}$ evaluated at $M_Z$ and the specific nucleon decay mode.
\item $LF$ is a {\bf Loop Factor} depending on gaugino, squark and slepton masses roughly as $M_{(1/2)}/m_0^2$ for $M_{(1/2)} << m_0$.
\end{itemize}

The Flavor Factor (FF) is model dependent.  In our analysis we use a particular SO(10) SUSY GUT with a $U(2)\times U(1)^n$ family symmetry with the Yukawa matrices given in \cite{brt}.  Note, however, that the Yukawa matrices are fixed to fit the low energy quark and lepton masses and mixing angles and the rate for $b \rightarrow s \gamma$. The model is also constrained by electroweak symmetry breaking.   We estimate, by comparing two quite different models, that the model dependence is probably no more than an order of magnitude in the rate.

The largest uncertainties in the proton lifetime, however, enter through the value of the effective color triplet mass $\mt$; through the sparticle spectrum via the Loop Factor (LF), and through the strong interaction matrix element $\beta_{lat}$.   We address these uncertainties below.

\begin{center}
{\bf Effective Color Triplet Mass -  \mt}
\end{center}

The color triplets are required to be heavy with mass of order $M_G$ since they contribute to nucleon decay;  Higgs doublets on the other hand must have mass of order the weak scale.  In SO(10) the Higgs doublets ($H_u, \; H_d$) and triplets ($T, \; \bar T$) are in the field $10_H$ which is the only Higgs-like representation coupling to standard model fermions.  A simple mechanism for accomplishing this doublet-triplet splitting exists in SO(10), known as the Dimopoulos-Wilczek [DW] mechanism\cite{DW}.  In this mechanism an adjoint (45 dimensional) scalar obtains a vev of order (B - L) $\times M_G$, with B (Baryon) and L (Lepton) number and gives mass to the Higgs field $10_H$.  Since color triplet Higgs fields have non-zero B-L charge, while Higgs doublets have zero charge, only the triplets obtain mass.   A simple variation of the original mechanism also allows for the possibility of obtaining large (or small) $\tan\beta \sim 50 (2)$ solutions with the addition of four fields, $\bar \psi^\prime, \; \bar \psi $; $\psi^\prime, \; \psi$ ($ \overline{16}, \; 16 $ representations of SO(10)) with the unprimed fields obtaining vevs in the "right-handed neutrino" directions.  For self-consistency,   $\psi, \; \bar \psi$ get mass of order $M_G$ and their mass and vevs are generated in the SO(10) breaking sector of the theory\cite{mohapatra}. 

The Higgs doublet and triplet mass matrices are given by

\hspace{1.8in} $\begin{array}{ccc} 10_H^5 & \; 10^5 & \; (\bar \psi^\prime)^5  \end{array}$ 
\vspace{-.1in}
\begin{equation} M_{(t,\ d)} = \left(\begin{array}{ccc} 
0 & \langle 45 \rangle & 0 \\
 \langle 45 \rangle & X & \langle \psi \rangle  \\
\langle \bar \psi \rangle & 0  & M \end{array}\right) \left(\begin{array}{c} 
10_H^{\bar 5} \\ 10^{\bar 5} \\ (\psi^\prime)^{\bar 5} \end{array}\right) \nn
\end{equation}
where the superscript indicates the SU(5) content of the field.  We take the vevs  $\langle 45 \rangle \sim (B - L) M_G$,  $M \sim M_G$ and $X \sim 10^{-3} \; M_G$.  We then consider two cases \begin{enumerate}
\item $\langle \bar \psi \rangle \approx \langle \psi \rangle = 0$
\item  $\langle \bar \psi \rangle \approx \langle \psi \rangle \sim 0.1 M_G$ 
\end{enumerate}

In both cases 1 and 2, the effective color triplet mass (eqn. 2) is given by
\begin{equation}
1/\mt \equiv (M^{-1}_t)_{11}=X M/det M_t \sim X/ M_G^2 \sim (10^{19} \;\; {\rm GeV})^{-1}
\end{equation}
where $M_t$ is the color triplet mass matrix.
Note there is actually no color triplet with mass greater than
$M_{Planck}$~\cite{babubarr}.

Consider the light Higgs doublets.

\begin{enumerate}
\item $H_u, \;\; H_d \subset 10_H$ are identified as the light Higgs doublets.  We have $\lambda_b = \lambda_\tau = \lambda_t \approx \lambda$ at $M_G$ and $\tan\beta \sim 50$.  $\lambda$ is the universal third generation Yukawa coupling given in $\lambda \; 10_H \; 16_3 \; 16_3$.
\item  The light Higgs doublets are identified as $H_u$ and $\gamma \ H_d$ with $\gamma =  \frac{X M}{\langle \bar \psi \rangle \; \langle \psi \rangle} << 1$.
We then have $\lambda_b = \lambda_\tau \approx \gamma \; \lambda << \lambda_t \approx \lambda$. 
\end{enumerate}
The lightest Higgs doublet (besides $H_u, \, H_d$) has mass of order $10^{-2} \, M_G$.

A limit on the value of $\mt$, the effective color triplet Higgs mass, is obtained by requiring perturbative threshold corrections to gauge coupling unification.   
At one loop the definition of the GUT scale is somewhat arbitrary.  A particularly convenient choice is to define $M_G$ as the scale where the two gauge couplings, $\alpha_i, \;\; i = 1,2$, meet.  We define $\tilde\alpha_G \equiv \alpha_1(M_G) =
\alpha_2(M_G)$ and the relative shift in $\alpha_3(M_G)$ is given by \begin{equation}\epsilon_3 \equiv  (\alpha_3(M_G) - \tilde\alpha_G)/\tilde\alpha_G . \end{equation} In general, a value of $\epsilon_3 \sim - (2 - 4\%)$ is needed to obtain $\alpha_s \sim \; 0.119$.
The one loop threshold correction coming solely from the Higgs sector ($5 + \bar 5$) is given by\cite{lucas2}
\begin{eqnarray}
 {\epsilon}_3(Higgs) = & {3 \tilde\alpha_G \over 5\pi}\log(\frac{\mt}{M_G}\gamma) &
\end{eqnarray}
which is valid in either case 1 or 2 above, if we let $\gamma = \frac{\lambda_b}{\lambda_t}$. 

Since the Higgs contribution to $\epsilon_3$ is always positive, we must therefore have a negative contribution coming from the rest of the GUT sector of the theory.  If we demand that the maximum allowed threshold correction
from the GUT sector is $- 10\% \;\; (- 8\%)$, we then have at most a positive $6\% \;\;(4\%)$ contribution from the Higgs sector (assuming we need $\epsilon_3 \sim -4\%$).  This gives an upper bound on the allowed values of $\mt \gamma$.

Note, in the small $\tan\beta$ regime there is no explicit suppression factor entering the coefficient of the dimension 5 operators, since they are all proportional to $\lambda^2/\mt$.  The difference in the small vs large $\tan\beta$ regimes is the maximum allowed value of $\mt$ consistent with perturbative threshold corrections. \footnote{This result was also discussed in \cite{bpw}.}  We find the bound \mt $ \ < 8 \times 10^{19} \gamma^{-1} \ {\rm GeV} \;\; (6 \times 10^{18} \gamma^{-1})$ GeV  for $\epsilon_3(Higgs) < $ 6\% (4\%).  Unfortunately this bound is exponentially sensitive to the assumed maximum allowed correction $\epsilon_3(Higgs)$.  

Finally we warn the reader that, with additional SO(10) adjoints and a clever modification of the DW doublet-triplet splitting sector, it is indeed possible to suppress proton decay via dimension five operators entirely, see for example Chacko and Mohapatra\cite{mohapatra}.  In this case, we would unfortunately lose a significant test of SUSY GUTs.  Moreover such elaborate constructs for suppressing nucleon decay seem entirely contrived and unnatural.

\vspace{.25in}
\begin{center}
{\bf Natural Superparticle Spectrum}
\end{center}

In order for SUSY to provide a solution to the gauge hierarchy problem,
the SUSY breaking scale $\Lambda_{SUSY}$ must be of order the weak scale.  Otherwise we must fine tune in order to have $M_Z \sim m_{Higgs} << \Lambda_{SUSY}$. In order for nucleon decay rates to be consistent with the present data we need to maximize squark and slepton masses, consistent with naturalness, and minimize gaugino masses, consistent with present experimental bounds.   

How heavy can squarks and sleptons be?  Since the first two families of squarks and sleptons couple weakly to the Higgs bosons, it has been argued that it is still natural to have heavy first and second generation squarks and sleptons as long as the third generation squarks and sleptons are lighter than, say, one TeV\cite{ckn}.
In fact in SUSY SO(10) with Yukawa unification and SO(10) boundary conditions at $M_G$ for soft SUSY breaking mass parameters, it was noted that the third generation squarks and sleptons are naturally lighter than the first two generations due to RGE running\cite{bagger,bcrw}.  This is due to the fact that in this limit, $\lambda_b = \lambda_\tau = \lambda_t \sim 1$ has the effect of driving the third generation scalars to lower masses.  For $m_0 = 3000$ GeV and large $\tan\beta$ we find all third generation squarks and sleptons are lighter than 1 TeV, {\it except} for the left-handed stau and tau sneutrino.  These have mass $\sim 2$ TeV.  If we estimate the contribution to $\delta m_H^2 \sim \frac{\lambda_\tau^2}{16 \pi^2} (\tilde m_\tau^2 + \tilde m_{\nu_\tau}^2)$ and demand $\delta m_H^2 < (130 \; {\rm GeV})^2$, we find $\tilde m_\tau \approx \tilde m_{\nu_\tau} < 2100$ GeV.  Thus we avoid fine-tuning in the effective theory at the weak scale.   

Note however that for $m_0 = 3000$ GeV we must still do some fine-tuning in order to obtain the correct electroweak symmetry breaking via RGE running from $M_{GUT}$ to $M_Z$.  This is because all three families of squarks and sleptons have large mass during most of the running.   This fine tuning can be avoided, however, if we take large masses at $M_G$ for the first two families of squarks and sleptons, while keeping the mass of the third family less than $\sim$ 1 TeV. This is certainly consistent with the SO(10) GUT $\times$ U (2) family symmetry which we are considering.  It may even be possible in this case to increase the squark and slepton masses of the first two families above 3 TeV.\footnote{In order to check these possibilities, we must include two loop RGE running as emphasized by Arkani-Hamed and Murayama~\cite{bagger}.  We will investigate this further elsewhere.}
 
 For small $\tan\beta$, only the stop squark mass is naturally light.  Thus, in this case, the upper bound on $m_0$ is $\sim$ 1000 GeV.

\vspace{.25in}
\begin{center}
{\bf Lattice Results}
\end{center} 
There have been several lattice calculations of the chiral Lagrangian parameters $\alpha_{lat}, \; \beta_{lat}$\cite{chadha,beta}.  A recent lattice calculation \cite{jlqcd} on a significantly larger lattice gives $\beta_{lat} \approx -\alpha_{lat} = 0.015$ GeV$^3$.   The statistical uncertainties in this result are small ($\pm 1$ in the last digit).  However systematic uncertainties connected with the chiral Lagrangian approach and the quenched approximation may be significant, perhaps as large as 50\%.  Note that the proton lifetime is also sensitive to the relative magnitude of $\alpha_{lat}$ and $\beta_{lat}$.  In our results we use the central value quoted in \cite{jlqcd}.

\section{Results}

The process $p \longrightarrow K^+ \bar{\nu}$ is the dominant decay mode for the proton.  In all cases we find the rate for $n \rightarrow K^0 \bar \nu$ dominates over $p \rightarrow K^+ \bar \nu$; typically by a factor of 2 - 4.   However the best experimental bound is on the latter ---
SuperKamiokande 90\% CL bounds on nucleon decay based on 61-ktonyear exposure\cite{superkbounds}  
\begin{eqnarray}
\tau(p \rightarrow K^+ \bar \nu) > & 1.9 \times 10^{33} \;\;{\rm years} &  \label{eq:superk}
\end{eqnarray}
and we use it to establish that SUSY GUTs are {\em under siege}.

The result of this analysis is the theoretical upper bound on the proton lifetime given by
\begin{eqnarray}
\tau(p \rightarrow K^+ \bar \nu) = & 4.7 \times 10^{33} \;\;{\rm years} \times (\frac{0.015 \; {\rm GeV}^3}{\beta_{lat}})^2 \times (\frac{\mt}{8 \times 10^{19} \;\; {\rm GeV}})^2 &  \label{eq:bound1}
\end{eqnarray}
for $m_0 = 3000$ GeV,  $M_{(1/2)} = 175$ GeV and $\tan\beta \sim 54$ or  
\begin{eqnarray}
\tau(p \rightarrow K^+ \bar \nu) = & 1.0 \times 10^{34} \;\;{\rm years} \times (\frac{0.015 \; {\rm GeV}^3}{\beta_{lat}})^2 \times (\frac{\mt}{5 \times 10^{21} \;\; {\rm GeV}})^2 &  \label{eq:bound2}
\end{eqnarray}
for $m_0 = 1000$ GeV,  $M_{(1/2)} = 300$ GeV and $\tan\beta = 2$.\footnote{For $\epsilon_3(Higgs) \le 4\%$ simply change the value of $\mt$ in eqn. \ref{eq:bound1} (\ref{eq:bound2})  to $6\times 10^{18} (4\times 10^{20})$.}  

Let us briefly summarize the theoretical input to these upper bounds.  
\begin{itemize}
\item  This upper bound assumes a very conservative upper limit on the color triplet Higgsino mass $\mt$; obtained by demanding that the Higgs contribution to the one loop threshold correction to gauge coupling unification at $M_G$ is at most 6\%.  

\item  Gaugino masses are taken to be near the allowed experimental lower bounds.   While squark and slepton masses must be near the upper bounds allowed by naturalness.  In SO(10) with universal squark and slepton mass $m_0$, we have $m_0 \sim 3000$ GeV for $\tan\beta \sim 50$ or $m_0 \sim 1000$ GeV for $\tan\beta \sim 2$.  

\item  We use the central value for $\beta_{lat}$ given in \cite{jlqcd}.
If the reader prefers another value, the new bound can be obtained by a simple rescaling.

\item   Any further uncertainty depends on the specific model for the Higgs doublet and triplet Yukawa couplings;  we estimate this uncertainty to be at most an order of magnitude in the lifetime.   The upper bound given here is for the particular SO(10) SUSY GUT with a $U(2)\times U(1)^n$ family symmetry which provides an excellent fit to charged fermion masses and mixing angles \cite{brt}.

\end{itemize}

Clearly we have pushed most of the parameters to (or perhaps beyond) what the reader may consider reasonable upper or lower bounds.    Nevertheless, with these exceptionally conservative bounds we are barely consistent with the latest Super-Kamiokande limits on $p \rightarrow K^+ \bar \nu$ (eqn. \ref{eq:superk}) \cite{superkbounds}.  

We have shown that the recent SuperKamiokande bounds on proton decay severely constrain SO(10) SUSY GUTs.  Recall, simple SU(5) SUSY GUTs have already been excluded by this data\cite{su5}.  Some general conclusions may be drawn from our analysis.  
\begin{itemize}
\item  Based on these results {\em the first two generations of squarks and sleptons must have mass significantly greater than 1 TeV}; while the third generation squarks and sleptons can be lighter than 1 TeV, {\em but not lighter than $\sim 400$ GeV}.  Thus, at best we expect only the third generation squarks and sleptons to be visible at LHC.
\item  Clearly proton decay must be seen soon IF minimal SUSY GUTs are the correct description of nature.
\end{itemize}

\section{Acknowledgments}
 
This work is supported in part by DOE grant  DOE/ER/01545-7895.  We are sincerely indebted to T. Bla\v zek for allowing us to use his computer code for the global $\chi^2$ analysis of the precision electroweak data including fermion masses and mixing angles.  SR thanks J. Pati and F. Wilczek for discussions.  Also RD thanks H. Murayama for discussions.

 \end{document}